# Autonomous Nonlinear Passive Transmit-Receive Switch for Compact IoT Devices: A Three-Port Agile Network

Rasool Keshavarz, Amber Abdullah, and Negin Shariati

RFCT Lab, University of Technology Sydney, Australia

{Rasool.keshavarz, Negin.shariati}@uts.edu.au, Amber.abdullah@student.uts.edu.au

*Abstract* — Recent advancements in RF technologies, especially Internet of Things (IoT) devices, require compact and integrated RF circulators or transmit-receive (T/R) switches for efficient resource use. Although conventional techniques are crucial in managing signal flow to prevent signal interference to sensitive receiver components, they have some drawbacks, such as limited isolation, low switching speed, complex circuitry, bulkiness, and high cost. This work presents a smart, miniaturized, nonlinear T/R switch capable of operating over a wide frequency range (0.8~1.3 GHz), making it suitable for IoT frequency bands. The switch achieves high isolation, low insertion loss, and intelligent transitions between transmitter and receiver without requiring external control or bias pins.

## I. INTRODUCTION

IoT devices utilize Transceiver (T/R) modules with circulators at their core for numerous applications, spanning from smart home automation to industrial monitoring and environmental sensing. These devices facilitate seamless communication, enabling remote control, data collection, and analysis, thereby enhancing efficiency, convenience, and decision-making across various domains [1, 2]. Circulators play a critical role in enabling two-way communication and Radar systems by controlling the flow of RF signals and providing isolation between transmit and receive paths. Although they provide high isolation in a wide frequency range and require no control circuits, they comprise ferrite materials to achieve non-reciprocity. This makes them costly and bulky, which can limit their use in compact and weight-sensitive IoT applications. Recently, magnet-less circulators have been presented to avoid the challenge of bulkiness [3, 4]. These three and four-port devices are based on the concept of spatiotemporal modulation and show nonreciprocal behavior to achieve controlled routing of RF signals and isolation. However, these circuits necessitate complex AC biasing circuits, including inductors, posing implementation challenges. Additionally, they exhibit narrow bandwidth and are unsuitable for compact IoT applications [5]. An alternate way to share the antenna is using a transmit-receive switch that alternates the connection between transmitter and receiver to control the signal flow and provide isolation [6]. However, this process requires precise synchronization to avoid conflicts where the transmitter and receiver try to access the antenna simultaneously and thus needs exact knowledge of Tx and Rx modes, information bits and bias lines. Although the T/R switches are compact and cost-effective, they come with the drawback of active switching circuitry, which introduces latency and a fixed response time that can affect the rapid switching speeds of the system and thus degrade the overall performance.

Therefore, a smart technique for fast switching is highly desirable, wherein the controller does not need to process information about the two modes or require any additional control circuitry. Additionally, integrated switch designs that can meet the space constraints of IoT technologies, with improved critical parameters, such as high isolation, and miniaturized footprint, are highly desirable. Conventionally, the reported passive T/R switches offer low bandwidth of only around ±10% of the center frequency due to narrow-band impedance transformation [6-8].

This work combines the functionalities of magnet-less circulators and T/R switches to introduce a new nonlinear autonomous passive switch with enhanced bandwidth, operating from 0.8~1.3 GHz. It achieves high isolation and rapid response in a compact size, requiring no control pins or biasing circuits. This three-port agile network leverages diode nonlinearity to transition the antenna between transmit and receive paths based on the incoming signal's power level. When the signal power level is high at the transmit port, the T/R switch routes the high-power signal from the transmitter to the antenna, placing the system in transmit mode. Conversely, in receive mode, when the power level is low at the antenna port, the switch redirects the antenna to the receiver port, capturing the low-power signals. Due to the low levels of IoT signals, this system employs Schottky diodes to achieve a minimal turn-on threshold voltage, thus enabling a low power level to enter Tx mode. As a result, the proposed design accomplishes a very high switching speed, low insertion loss, high isolation and is suitable for applications with extreme space constraints.

Section II of the paper covers the design of the proposed work. Section III compares the simulated/measured results with recent works. Lastly, Section IV summarizes the conclusion.

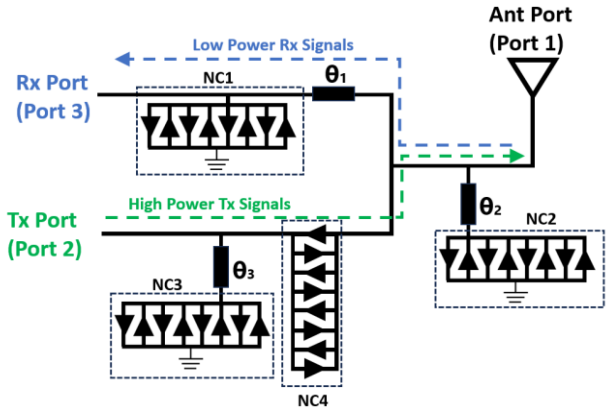

Fig. 1. Configuration of the proposed T/R system

## II. DESIGN AND WORKING PRINCIPLE

This section outlines the design process of the proposed autonomous T/R switch and provides a theoretical analysis of its Tx/Rx modes.

Fig. 1 shows the schematic of the proposed passive T/R switch. It consists of a three-port passive network comprising three impedance transformers ($IT_1, IT_2, IT_3$) and four nonlinear circuits ($NC_1, NC_2, NC_3, NC_4$) consisting of eight Schottky diodes each. The diodes are connected in an antiparallel configuration to allow RF signal flow in both directions. Additionally, the number of diodes is chosen appropriately to increase the nonlinearity and achieve highly sensitive operation to the incident power levels. The switch works in two modes, either transmit or receive mode, depending on the signal power level on all three ports. Fig. 2 below depicts the equivalent power-dependent nonlinear model of the Schottky diode, including package parasitic inductance and capacitance [9]. Moreover, Fig. 3 presents the real and imaginary parts of the grounded NC (nonlinear circuit). According to this figure, the input impedance of the NC can be determined for different incident power levels and over a wide frequency range (0.6~1.5 GHz) and the extracted values can be used in the design procedure.

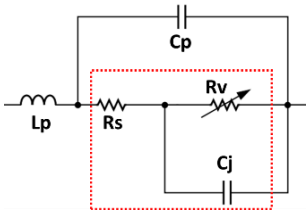

Fig. 2. Equivalent circuit model of a packaged Schottky diode

In transmit mode, the power level is high at the Tx port and $NC_4$ has very low impedance, which can be considered as a short circuit. Additionally, $NC_1$ shows low impedance, and power will be delivered to the Tx port under good matching condition and low insertion loss between Tx and antenna ports.

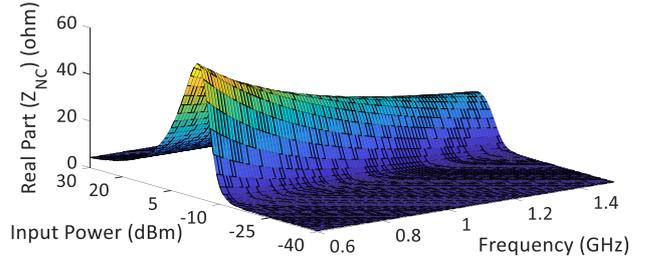

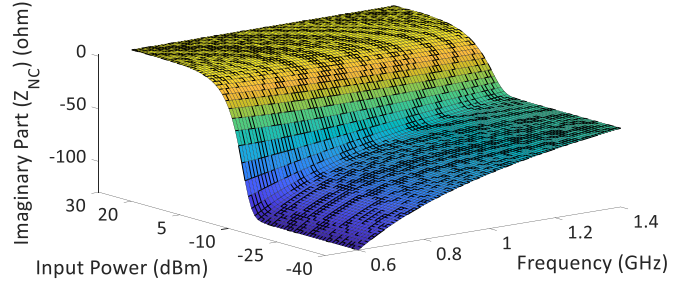

Fig. 3. Simulated results: (a) Real and (b) Imaginary, parts of the grounded NC sections in the proposed T/R switch for different power levels (-40 to +30 dBm) and over a wide frequency range (0.6~1.5 GHz).

The reflection and transmission coefficients between the antenna (port 1) and Tx (port 2), $S_{11}$ and $S_{21}$, are given by:

$$S_{11} = \frac{-Y_t Z_p}{2+Y_t Z_p}, \quad S_{21} = \frac{2}{2+Y_t Z_p} \quad (1)$$

Where

$$Y_t = Y_1 + Y_2 + Y_3 \quad (2)$$

and

$$\begin{cases} Y_1 = \frac{Z_{IT_1}+j(Z_{NC_1}+Z_p)\tan(\theta_1)}{Z_{IT_1}(Z_{NC_1}+Z_p+jZ_{IT_1}\tan(\theta_1))} \\ Y_2 = \frac{Z_{IT_2}+jZ_{NC_2}\tan(\theta_2)}{Z_{IT_2}(Z_{NC_2}+jZ_{IT_2}\tan(\theta_2))} \\ Y_3 = \frac{Z_{IT_3}+jZ_{NC_3}\tan(\theta_3)}{Z_{IT_3}(Z_{NC_3}+jZ_{IT_3}\tan(\theta_3))} \end{cases} \quad (3)$$

$Z_{IT_i}$ and $\theta_i$ are characteristic impedance and electrical length of *IT* lines, respectively, for *i=1,2,3*. Moreover, $Z_{NC_i}$ is input impedance of NC (*i=1,2,3*) and can be determined using Fig. 3 for the desired frequency and power level. $Z_p$ is port impedance and is considered 50 Ω in the design process.

During receive mode, a low-power desirable signal is present at the antenna port, while no signal is present at the Tx port. Due to the low power level of the input signal, $NC_4$ and $NC_1$ show high impedances and can be considered as open circuits. Hence, the reflection and transmission coefficients between antenna and Rx ports, $S_{11}$ and $S_{31}$, are given by:

$$S_{11} = \cos(\theta_1) + jZ_{IT_1}Y_2\sin(\theta_1), \quad S_{21} = jZ_{IT_1}\sin(\theta_1) \quad (4)$$

Due to the complexity of closed-form modeling for IT lines and NCs, as well as the presence of parasitic components in the proposed TR switch, we employed a Genetic Algorithm in

Advanced Design System (ADS) software. By utilizing the EM/circuit co-simulation technique, we determined the proper lengths and characteristic impedances of the IT lines and T/R switch is realized. In Section III, simulated and measured results for a specific case will be presented and discussed.

## III. SIMULATION AND MEASUREMENT RESULTS

The proposed T/R switch is designed, simulated and fabricated on FR4 substrate with thickness of 1.6 mm, $\tan\delta = 0.03$ and $\varepsilon_r = 4.6$ (Fig. 4). The structure is simulated in ADS using Harmonic Balance and Large Signal S-parameter simulations. The SPICE parameters of the Schottky diode SMS7621 are used to characterize the nonlinear analysis of the switch, along with its package parasitic series inductance $L_P$ and parallel capacitance $C_P$. A parametric optimization simulation was conducted using the Genetic Algorithm for input power ranging from -40 dBm to +30 dBm over the frequency range of 0.8~1.3 GHz to assess the performance of the switch in both receive and transmit modes. The designed parameters of the T/R switch and diode spice model are presented in Table 1.

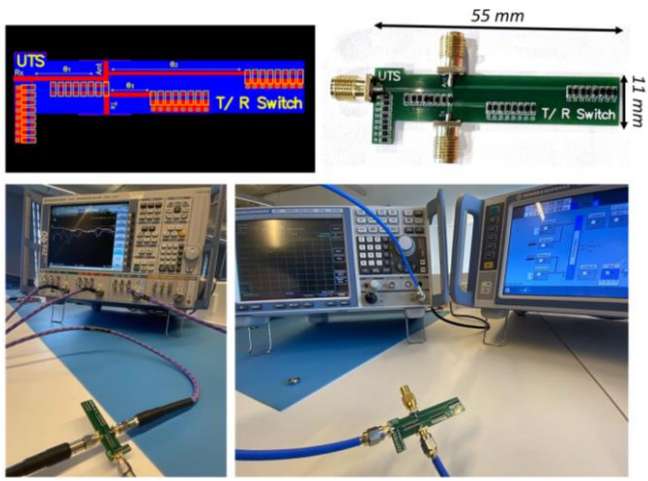

Fig. 4. Photographs of the: PCB layout, fabricated prototype; and testing setup using VNA, signal generator and spectrum analyzer.

Table 1. Values of the designed parameters.

| IT Lines Parameters | $\theta_1$ | $\theta_2$ | $\theta_3$ | $Z_{IT_1}$ | $Z_{IT_2}$ | $Z_{IT_3}$ |
|---|---|---|---|---|---|---|
| Value | $28^0$ | $86^0$ | $25^0$ | $89\Omega$ | $97\Omega$ | $84\Omega$ |
| Diode Parameters | $R_s$ | $C_{j0}$ | $L_P$ | $C_P$ | $V_j$ | M |
| Value | $12\Omega$ | 0.1pF | 2nH | 0.08pF | 0.5V | 0.35 |

Fig. 5 shows measured |S|-parameters of the proposed T/R switch at low power range (-30 dBm). According to this figure, the insertion loss between the antenna and Rx port is less than 1 dB and isolation between Tx and Rx ports is higher than 18 dB at low power range. In addition, this figure exhibits return loss better than 10 dB over the wide frequency range of 0.8~1.3 GHz.

To evaluate the power-dependent-nonlinear-passive operation of the switch at 1.2 GHz, a power sweep from -40 to +30 dBm is applied at the Ant port, and the power levels at the Rx and Tx ports are measured. The results are shown in Fig. 6 (a). It is depicted that for low power levels (<-5dBm), the Rx mode is turned on, and the power is transmitted to the Rx port, with a small insertion loss, while there is a high loss at the Tx port. Meanwhile, the switch converts to the Tx mode for high power levels (>+20 dBm), and the power is transmitted to the Tx port with high isolation from the Rx port.

Fig. 6 (b) exhibits insertion loss of the T/R switch in two operating modes. The insertion loss in the Rx mode is better than 0.7 dB for low power range, <-10 dBm. So, the proposed switch can be used in low noise figure receiver chains for highly sensitive receivers. In addition, insertion loss of T/R switch in Tx mode is less than 1 dB. Moreover, the isolations between Tx and Rx ports are higher than 17 dB and 15 dB in Rx and Tx modes, respectively (Fig. 6 (c)).

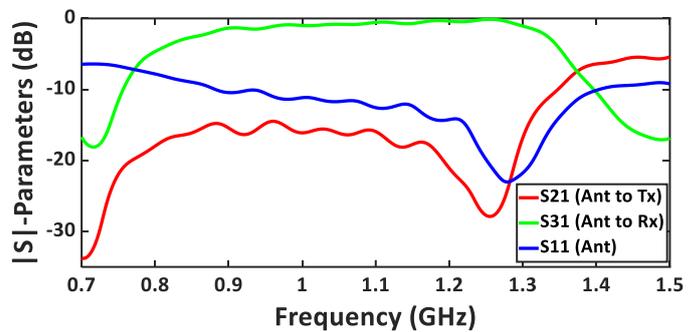

Fig. 5. Measured |S|-Parameters of the proposed T/R switch (low power, -30 dBm).

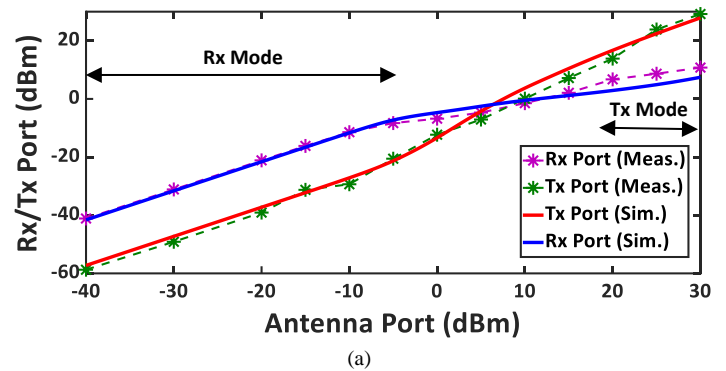

(a)

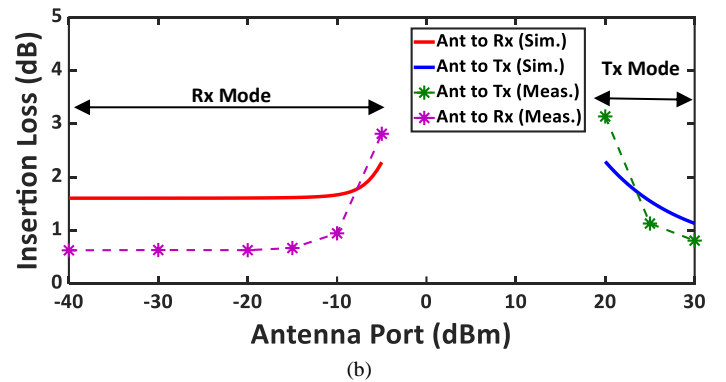

(b)

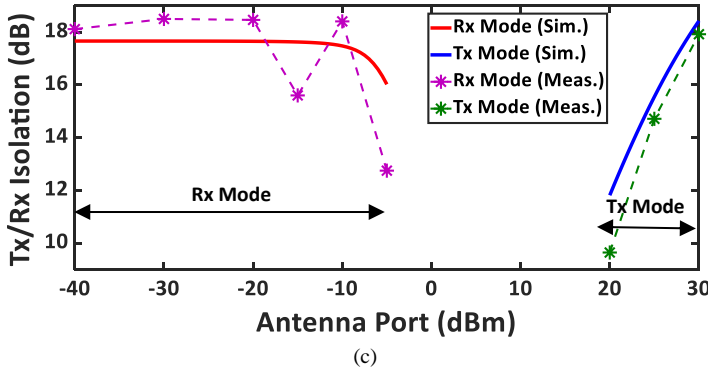

Fig. 6. Simulated and measured results: (a) output power levels at Tx and Rx ports, (b) antenna to Rx/Tx insertion loss, and (c) isolation b/w Tx and Rx at 1.2 GHz for the antenna power of -40 to +30 dBm.

The presented work is compared with other recent relevant works, as shown in Table 2. The proposed design, unlike other most of the T/R switches and magnet-less circulators, does not require AC or DC bias. This feature enhances integration, speed, and compactness. The presented structure achieves a comparable insertion loss of 0.6 dB and a high 18 dB isolation bandwidth of 45% with a compact footprint of 55mm x 11 mm over the 0.8~1.3 GHz range, which is smaller than other works in terms of electrical length λ. From the discussion above, it can be concluded that the proposed research work is highly suitable for compact IoT applications requiring high isolation bandwidth and low insertion loss.

Table 2. Comparison of the presented work with other published works.

| Metrics | [3] | [4] | [5] | [11] | This Work |
|---|---|---|---|---|---|
| Bias | AC | AC, DC | AC, DC | AC, DC | No |
| Isolation (dB) | 18 | 20 | 20 | 30 | 18 |
| Center Freq (GHz) | 1 | 1 | 1 | 1 | 1 |
| Bandwidth (MHz) | 35 | 40 | 20 | 10 | 500 |
| Size (mm) | N/A | N/A | 13x11* | N/A | 55x11 |
| Insertion Loss Ant-Rx (dB) | 3.4 | 2 | 2 | 3 | 0.6 |

*Excluding AC and DC biasing networks.

## IV. CONCLUSION

A smart, low-cost, low-loss, passive, and compact T/R switch is presented, which operates based on the nonlinear behavior of diodes. The proposed T/R switch is designed, simulated, fabricated, and measured to validate the results. It operates in Rx mode at low power and switches to Tx mode at high power. The switch provides high isolation of 15 dB in Tx mode and 18 dB in Rx mode, with an insertion loss of less than 0.6 dB in Rx mode. This addresses the issues of bulkiness in circulators and eliminates the need for AC/DC biasing in magnet-less circulators and T/R switches.

ACKNOWLEDGMENT

This research was supported by funding from NTT Group (Nippon Telegraph and Telephone Corporation) and Food Agility Cooperative Research Centre (CRC) Ltd, funded under the Commonwealth Government CRC Program. The CRC Program supports industry-led collaborations between industry, researchers and the community.